\documentclass[preprint,3p,times]{elsarticle}

\usepackage[T1]{fontenc}
\usepackage[utf8]{inputenc}

\usepackage{pgf}
\usepackage{graphicx}

\usepackage{amsmath}
\usepackage{amssymb}

\usepackage{lineno}

\biboptions{sort&compress}

\usepackage{algorithm}
\usepackage{algorithmicx}
\usepackage{algpseudocode}
\usepackage{array} 
\usepackage{booktabs} 
\usepackage{hyperref}
\usepackage{multirow}
\usepackage{rotating}
\usepackage[expproduct = \cdot]{siunitx}
\usepackage{tikz}
\usetikzlibrary{arrows,shapes}
\usepackage{xcolor}

\journal{Computational Statistics \& Data Analysis }

\newcommand{\nth}[1]{#1\textsuperscript{th}}
\newcommand{\refalg}[1]{Algorithm~\ref{#1}}

\newcommand{\reffig}[1]{Fig.~\ref{#1}}
\newcommand{\refsec}[1]{Section~\ref{#1}}
\newcommand{\reftab}[1]{Table~\ref{#1}}

\newif\iftodo\todofalse

\newif\ifpgf\pgffalse

\newcommand{\doubbar}{\,||\,}
\renewcommand{\|}{\ensuremath{\,|\,}}
\DeclareMathOperator{\diag}{diag}
\DeclareMathOperator{\ESSop}{ESS}
\newcommand{\ess}{\ensuremath{\ESSop}}
\newcommand{\estZ}{\ensuremath{\widehat{Z}}}
\newcommand{\estDelZ}{\ensuremath{\widehat{\Delta Z}}}
\DeclareMathOperator{\LogGamma}{LogGamma}
\newcommand{\kl}{\ensuremath{\mathrm{KL}}}
\newcommand{\matsig}{\ensuremath{\boldsymbol{\Sigma}}}
\newcommand{\order}[1]{\mathcal{O}\left({#1}\right)}
\newcommand{\perpl}{\ensuremath{\mathcal{P}}}
\newcommand{\rmdx}[1]{\mbox{d} #1 \,} 
\DeclareMathOperator{\shell}{circ}

\newcommand{\vecc}{\ensuremath{\boldsymbol{c}}}
\newcommand{\vecmu}{\ensuremath{\boldsymbol{\mu}}}
\newcommand{\vecth}{\ensuremath{\boldsymbol{\theta}}}

\newlength{\relwidth}

\newcommand\tabvsptop{\rule{0pt}{2.6ex}}

\begin{document}

\begin{frontmatter}



\title{Initializing adaptive importance sampling with Markov chains}


\author{Frederik Beaujean\corref{cor1}}
\cortext[cor1]{Corresponding author}
\ead{beaujean@mpp.mpg.de}

\author{Allen Caldwell}
\ead{caldwell@mpp.mpg.de}

\address{
Max Planck Institute for Physics
}

\begin{abstract}
  Adaptive importance sampling is a powerful tool to sample from complicated
  target densities, but its success depends sensitively on the initial proposal
  density. An algorithm is presented to automatically perform the initialization
  using Markov chains and hierarchical clustering. The performance is checked on
  challenging multimodal examples in up to 20 dimensions and compared to results
  from nested sampling. Our approach yields a proposal that leads to rapid
  convergence and accurate estimation of overall normalization and marginal
  distributions.

\end{abstract}

\begin{keyword}
  adaptive importance sampling \sep population Monte Carlo \sep Markov chain \sep hierarchical clustering
  \sep multimodal

\end{keyword}

\end{frontmatter}


\section{Introduction}\label{sec:introduction}

The fundamental problem we wish to address is to sample from or integrate over a
complicated density $P(\vecth)$, the \emph{target density}, in a moderately
high-dimensional parameter space. Our main application is Bayesian statistics,
where $P(\vecth)$ is identified with the unnormalized posterior
distribution. The samples are useful for parameter inference, and for model
comparison it is necessary to compute the posterior normalization---the evidence
or marginal likelihood, $Z$---given by the integral over the product of
likelihood $L(\vecth)$ and prior $P_0(\vecth)$ as
\begin{equation}
  \label{eq:def-evidence}
  Z =  \int \rmdx{\vecth} P(\vecth) = \int \rmdx{\vecth} L(\vecth) P_0(\vecth).
\end{equation}
A plethora of sampling algorithms exists in the literature; a comprehensive,
though somewhat dated, overview is presented in \cite{Casella:2004}. In
contrast to algorithms tailored to very specific targets, we rather want to
make progress toward the ``holy grail'' of sampling: an efficient, accurate, and
parallelizable algorithm that copes with any (continuous) target in any
dimension. Ideally, such an algorithm yields samples and the integral in one
run.

Current analyses at the frontier of cosmology and elementary particle physics
often involve extensions of the accepted standard models with a large number of
parameters, but only loose constraints from the available data; see, for
example, \cite{AbdusSalam:2010, Ade:2013lta}. The methods presented here were
developed in the course of a global analysis of rare B-meson decays described in
detail in~\cite{Beaujean:2012uj,Beaujean:2012}. Posterior densities occurring in
that analysis exhibit many of the typical features that require a sophisticated
sampling procedure as there are degeneracies due to continuous symmetries and
multiple isolated modes arising from discrete symmetries in $d=18-31$
dimensions. The standard approach based on local random-walk Markov chains
(MCMC) is notorious for failing to produce reliable estimates under these
circumstances as chains mix very slowly or not at all.

In challenging problems with large data sets or difficult-to-obtain model
predictions, the sampling is further complicated by the relatively long time
required to evaluate the target density (roughly \SI{1}{s} in the motivating
example~\cite{Beaujean:2012uj}). Apart from more efficient algorithms, the most
straightforward option to reduce the wallclock time needed for the computation
is to use parallel computing facilities. The easiest implementation is done on a
standard system with multiple cores, but it is desirable to use computing clusters
or graphical processing units offering hundreds or even thousands of cores.

Adaptive importance sampling, or population Monte Carlo (PMC)
\cite{Cappe:2004,cappe_adaptive_2008}, is an evolution of classic importance
sampling \cite{Neumann:1951,Casella:2004} combining most of the desirable
features outlined above. The basic idea is to use a mixture density as a
proposal function $q(\vecth)$, and to iteratively update $q(\vecth)$ to match
the target density as closely as possible. The individual components of
$q(\vecth)$ are conveniently chosen as multivariate Gaussian or Student's t
distributions; by adjusting their location and covariance one can easily
accommodate multiple modes and degeneracies. Each proposal update requires a
step where a large number of i.i.d. samples $\left\{ \vecth^i: i=1\dots N
\right\}$ are drawn from $q(\vecth)$, allowing trivial parallelization of the
potentially costly evaluation of the importance weights $w_i =
P(\vecth^i)/q(\vecth^i)$.

The PMC update algorithm is based on expectation-maximization (EM)
\cite{Dempster:1977} and seeks to reduce the Kullback-Leibler divergence
\cite{Kullback:1951} between target and proposal, $\kl(P \doubbar q)$. Since
$\kl(P \doubbar q)$ usually has multiple minima, and EM tends toward a local
minimum, the initial guess for the proposal is of utmost importance for the
success of the algorithm. With a poor proposal, PMC fails as proposal updates
lead to a consecutively poorer approximation of the target. In that case,
typically more and more of the proposal components are assigned vanishing weight
until only one remains. Since a single component is insufficient in all but the
most trivial cases, the PMC results are useless. The authors of PMC and its
reference implementation~\cite{pmclib:2011} in the C language commented on the
issue of the initial guess, but provided only basic advice that is useful just
in fairly simple unimodal problems. In their first applications of PMC to
physics analyses~\cite{Wraith:2009if, Kilbinger:2009by}, it was sufficient to
scatter mixture components around the center of the parameter range or around a
previously computed maximum using the Fisher matrix and ``educated
guesses''~\cite{Kilbinger:2009by}. 
However, this approach did not give satisfactory results for the analysis
presented in~\cite{Beaujean:2012uj}.

Clearly, a more robust approach to initialization is preferable.  Previous
attempts at such an initialization were suggested in the context of
econometrics~\cite{Hoogerheide:2011} and population
genetics~\cite{Cornuet:2012}. In the basic algorithm of \cite{Hoogerheide:2011},
the authors propose to start with a single component given by the mode and the
inverse Hessian at the mode. In each update, one new component is added, which
is constructed from the highest importance weights. The algorithm terminates
when the standard deviation of the weights divided by their mean is sufficiently
small. For multimodal problems, they suggest a tempering approach to first adapt
the proposal to a simplified target and demonstrate successful discovery of 20
well separated Gaussians, albeit only in 2D.  For nonelliptical target densities
in higher dimensions where many dozens of components are needed, their approach
would presumably require an excessive number of updates.
In \cite{Cornuet:2012}, a large sample from the uniform or prior distribution
with a logistic rescaling is used to learn the features of the target. The
authors propose to run Gaussian mixture clustering with the integrated
likelihood criterion fixing the optimal number of components of the initial
proposal for PMC{}. By cleverly combining the samples of \emph{all} PMC{} update
steps, and not only the most recent one as in our approach, they report a
significant Monte Carlo variance reduction.
%
Nested sampling~\cite{Skilling:2006} is an alternative technique to
simultaneously compute weighted samples and the normalization of a complicated
target density. The basic idea is to evolve a collection of sample points such
that in each iteration the point with the smallest likelihood is replaced by a
new point with larger likelihood drawn from the prior.
Multinest~\cite{Feroz:2008xx} is the most widely used implementation of nested
sampling; it uses sets of ellipsoids to map out the target's regions of
interest.

Our approach is similar to the efforts of~\cite{Hoogerheide:2011,Cornuet:2012}
in that we seek to create a good initial proposal for PMC with a minimum of
manual intervention. The initialization proceeds in two phases. In the learning
phase, multiple local random-walk Markov chains are run in parallel to discover
and explore the regions where $P$ is large. In the next phase, we use the chains
to extract the local features of the target by partitioning up chains into short
patches of around 100 iterations and thus turn one of the weaknesses of the
random walk---the slow diffusion-like exploration---into a virtue. Each patch
defines a mixture component through its sample mean and covariance. To reduce
the number of components to a tractable number, we employ hierarchical
clustering \cite{Goldberger:2004}. The proposed initialization differs from
previous attempts in its usage of Markov chains and is designed specifically for
complicated targets in fairly high dimensions $d \lesssim 40$ with complicated
shapes such as degeneracies, multiple modes, and other nonelliptical structures.

After a detailed description of the algorithm in \refsec{sec:algorithm}, and a
brief summary of its parameters in \refsec{sec:short-guide-param}, we illustrate
the algorithm with several examples in various dimensions in
\refsec{sec:examples} and compare its performance to that of Multinest version
2.18. We do not compare to alternative PMC initializations since no
implementations were available to us when this work was carried out.  Ideas on
future directions and concluding remarks are presented in
Sections~\ref{sec:outlook} and~\ref{sec:conclusion}.

\section{Algorithm}\label{sec:algorithm}

\subsection{Overview}\label{sec:overview}

Our focus is on creating a good initial proposal for PMC with a minimum of
manual intervention. In general, it is  necessary to
\begin{enumerate}
  \item explore the target $P$ through evaluations at a number of points in
  parameter space; and to
  \item  extract and combine the knowledge from the exploration into a
  mixture density that approximates $P$.
\end{enumerate}
In practice, our suggestion is to \emph{combine the best of MCMC and PMC} in
  three steps (cf. \reffig{fig:algo-flow}):

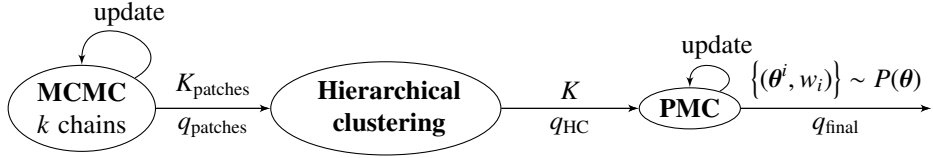
\begin{figure}[t]
  \begin{center}
    \tikzstyle{line} = [draw, -latex']
    \tikzstyle{cloud} = [draw, ellipse,fill=red!0, inner sep=2pt, node distance=40mm]
    \tikzstyle{empty} = [inner sep=0pt]

    \begin{tikzpicture}[auto, every loop/.style={in=90, out=30, looseness=3}]
      \node[cloud, text width=3.5em] (mcmc){
        \begin{minipage}[h]{\textwidth}
          \centering
          \textbf{MCMC}\\
          $k$ chains
        \end{minipage}
      };
      \node[cloud, right of=mcmc] (hc){
        \settowidth{\relwidth}{Hierarchicalx}
        \begin{minipage}[h]{\relwidth}
          \centering
          \textbf{Hierarchical\\
          clustering}
        \end{minipage}
      };
      \node[cloud, right of=hc] (pmc){
          \textbf{PMC}
      };
      \node[empty, node distance=32mm, right of=pmc] (out){};

      \path [line]
      (mcmc)
      edge  node [above] {$K_{\rm patches}$} (hc)
      edge  node [below] {$q_{\rm patches}$} (hc)
      edge [loop above] node {update} (mcmc)

      (hc) edge node [above] {$K$} (pmc)
      (hc) edge node [below] {$q_{\rm HC}$} (pmc)
      (pmc)
      edge [loop above] node {update} (pmc)
      edge node [above] {$\left\{ (\vecth^i, w_i) \right\} \sim P(\vecth)$} (out)
      edge node [below] {$q_{\rm final}$} (out);

    \end{tikzpicture}

  \end{center}
  \caption{\label{fig:algo-flow} Overview of the algorithm described in
    \refsec{sec:overview}. The different $q$ refer to the mixture densities used
    at different stages of the algorithm, while the $K$ refer to the number of
    components.}
\end{figure}

\begin{enumerate}
  \item We run $k$ Markov chains in parallel, each performing a local random walk
  with an adaptive proposal on the target $P(\vecth)$.
  \item Then we extract the support of the target density by splitting each
  chain into many small patches. Sample mean and covariance of each patch define
  one multivariate density, the collection of $K_{\rm patches}$ patches yields a
  mixture density $q_{\rm patches}(\vecth)$.
  \item Typically, there are more patches than actually needed; hierarchical
  clustering produces a mixture with far fewer components $K \ll K_{\rm patches}$ but
  essentially the same knowledge of the target density by removing redundant
  information. The output of hierarchical clustering, $q_{\rm HC}(\vecth)$, is
  used with minor modifications as the initial proposal for PMC.  We then run
  the standard PMC updates implemented in~\cite{pmclib:2011} until convergence
  and extract samples $\vecth^i$ with importance weights $w_i$ using the
  proposal $q_{\rm final}(\vecth)$.
\end{enumerate}

The combination of MCMC and PMC is one leap forward towards a black-box Monte
Carlo sampler that learns the relevant features of the target density
automatically.  In order to make optimal use of a parallel computing
infrastructure, the MCMC prerun is to be kept at a minimum length, and
preferably most evaluations of the target density are performed during the PMC
phase, when massive parallelization is available.

\subsection{Detailed description}\label{sec:detailed-description}

\subsubsection{MCMC prerun}\label{sec:mcmc-prerun}

For concreteness, we chose to implement the adaptive Metropolis algorithm
suggested in \cite{Haario:2001} for the examples discussed below. It uses a
multivariate Gaussian proposal that is centered on the current point and updated
based on the covariance of previous iterations with the cooling scheme described
in \cite{Wraith:2009if}. Note that the resulting chain strictly speaking is not
a Markov chain because the proposal is continuously adapted after every batch of
$N_{\rm update}$ iterations, but for simplicity, we continue to use the term
``Markov''. In our algorithm, we only rely on the fact that the samples are
generated by a local random walk and that their asymptotic distribution is
$P(\theta)$.

Assuming no knowledge of the target density other than that is zero outside of a
given hyperrectangle in $\mathbb{R}^d$, we draw the initial positions of the
chains from the uniform distribution. If the target is a posterior density and
the priors are of a simple form, we can draw the starting points directly from
the prior. Similarly, the initial covariance matrix is proportional to
\begin{equation}
  \label{eq:sigma-initial}
  \matsig^0 = \diag \left( \sigma_1^2, \sigma_2^2, \dots, \sigma_d^2 \right) ,
\end{equation}
where $\sigma_i^2$ is the prior variance of the \nth{i} parameter. We then
rescale $\matsig^0$ by $2.38^2/d$~\cite{Roberts:1997} to increase the efficiency
of the initial proposal; in subsequent updates, we update the scale factor to
achieve an acceptance rate between $(15-35)\,\%$.

In most problems, the prior is significantly more diffuse than the posterior,
hence our choice of seeding the chains automatically generates
overdispersion. The main reason why overdispersion is desirable to us is that in
the face of potentially several maxima, with little analytical knowledge of a
posterior that is often available only in the form of computer code, it is
imperative to explore the full parameter space and to find \emph{all} regions of
significant probability. These regions are not limited to local maxima, but
include degenerate regions as well. Therefore, the number of chains, $k$, should
be chosen significantly larger than the number of expected maxima. If the
location of the maxima is known, an equal number of chains can be started in
each maximum for higher efficiency. Since in many realistic problems the purpose
of the sampling is to discover the maxima, in the examples below we choose
\emph{not} to use the available analytical knowledge on the location of the
modes in order to highlight potential pitfalls.

We select a fixed number of iterations, $N_{\rm MCMC}$, to terminate the
sampling without regard for chain mixing. $N_{\rm MCMC}$ ought to be chosen
rather small in the trade-off between accuracy and computing time. Even in the
most complicated settings that we treated~\cite{Beaujean:2012}, $N_{\rm MCMC}
\lesssim \num{40000}$ revealed enough information about $P(\vecth)$.

Given the prerun of $k$ chains, we extract the local information by exploiting
the slow, diffusion-like exploration of the chain. To this end, we choose a
\emph{patch length} $L$, and partition the history of each chain, with the
exception of the burn-in, into patches of length $L$.  For the \nth{$i$} patch,
we compute the sample mean $\vecmu_i$ and sample covariance $\matsig_i$, and
form a multivariate Gaussian density. Patches in which no move is accepted are
discarded, and those for which the numerical Cholesky decomposition fails are
used with off-diagonal elements of the covariance matrix set to zero.  The patch
length ought to be chosen in such a way that small-scale features of the
posterior can be explored during $L$ iterations. A good value of $L$ slightly
increases with $d$ and possible degeneracies. On the other hand, $L$ must not be
too small, else the chain cannot move enough. Combing patches from all $k$
chains, we obtain a Gaussian mixture density of $K_{\rm patches}$ components
\begin{equation}
  \label{eq:patch-mixture}
  q_{\rm patches}(\vecth) \equiv \sum_{i=1}^{K_{\rm patches}} \alpha_i \mathcal{N}(\vecth \| \vecmu_i, \matsig_i) .
\end{equation}
and assign equal weight $\alpha_i= 1/K_{\rm patches}$ to each component. Note that we
do \emph{not} take into account the value of the posterior in each patch;
rather, we will ultimately rely on PMC to find the proper component weights.

\subsubsection{Hierarchical clustering}\label{sec:hier-clust}

The information about the target from the Markov chains is contained in a
mixture density with a large number of components $K_{\rm patches}$, where typical
values of $K_{\rm patches}$ can reach several thousands. For computational efficiency,
it is important to reduce the complexity of the mixture to keep the number of
samples needed in each PMC update step low. At the same time, we wish to
preserve as much information as possible. The goal is to compress the
$K_{\rm patches}$ components into a mixture with only $K \ll K_{\rm patches}$ components by
removing redundant information that, for example, comes from multiple chains
that mix or from a single chains repeatedly visiting a region.
\emph{Hierarchical clustering} \cite{Goldberger:2004} is our weapon of
choice. It achieves the compression by finding a $K$-component Gaussian mixture
\begin{equation}
  \label{eq:qHC}
  q_{\rm HC}(\vecth) \equiv \sum_{i=1}^{K} \alpha_i \mathcal{N}(\vecth \| \vecmu_i, \matsig_i) .
\end{equation}
that minimizes a distance measure based on the Kullback-Leibler divergence
\cite{Kullback:1951}.

\subsubsection*{Initialization}
\label{sec:hc-initialization}

\newcommand{\vecn}{\ensuremath{\boldsymbol{n}}}

Hierarchical clustering, being an expectation-maximization variant, converges
only on a local minimum of the distance measure. Given a large number of input
components, there exist numerous local minima, hence it is crucial to supply
good initial guesses for the output components, such that the
initial solution is already very close to a \emph{good} final solution. We then
note rapid convergence after $\order{10}$ steps.  There are two important
questions to address:
\begin{enumerate}
  \item Where to put the initial output components?
  \item How many output components, $K$, are needed?
\end{enumerate}
At present, we assume a fixed value of $K$.  In~\cite{Goldberger:2004}, it is
vaguely recommend to use ``standard methods for model selection'' to determine
$K$. We can only speculate that they refer to the Bayesian information criterion
\cite{Schwarz:1978} or the Akaike information criterion
\cite{Akaike:1974}. Another approach would be to add one component at a time
until $K$ is ``large enough''. It then remains to specify a quantitative
stopping criterion. In \cite{Hoogerheide:2011}, an attempt at such a criterion
is presented, but it appears somewhat inefficient when a large number of
components is needed. As a rule of thumb, we recommend $K$ should be at least as
large as $d$.

But to answer the first question, we have a good idea \emph{where} to place the
components. The key is to group the chains, and to have a fixed number of
components per group from \emph{long patches}; i.e., parts of chains of length
significantly exceeding $L$. To begin with, it is necessary to determine which
chains have mixed in the prerun. Two or more chains whose common Gelman-Rubin
$R$ values~\cite{Gelman:1992} are less than a given constant, $R_c$, for all
parameters, form a \emph{group} of chains. Most importantly, this ensures that a
similar and sufficient number of components is placed in every mode of the
target density, regardless of how many chains visited that mode. We ignore the
burn-in samples of each chain as described in the previous section.

Let us assume we want $K_g$ components from a group of $k_g$ chains. If $K_g
\geq k_g$, we find the minimal lexicographic \emph{integer partition} of $K_g$
into exactly $k_g$ parts. Hence, the partition, represented as a
$k_g$-dimensional vector of integers \vecn{}, is given by
\begin{equation}
  \label{eq:int-partition}
  \vecn =  \left( \left\lceil \frac{K_g}{k_g} \right\rceil, \dots , \left\lceil \frac{K_g}{k_g} \right\rceil,
  \left\lfloor \frac{K_g}{k_g} \right\rfloor, \dots , \left\lfloor \frac{K_g}{k_g} \right\rfloor \right) ,
\end{equation}
where we use the ceiling ($\lceil \rceil$) and floor ($\lfloor \rfloor$)
operations. The first $\left(K_g \mod k_g\right)$ parts are one larger than the
remaining parts. For example, with $K_g = 6$ and $k_g = 4$, the partition is
$\left( 2,2,1,1 \right)$.  In the rare case $K_g < k_g$, the integer
partitioning cannot be performed as above. Instead, we combine all individual
chains into one long chain, and set $k_g =1$.

Finally, the \nth{$i$} chain is partitioned into $n_i$ long patches, and the
sample mean and covariance of each patch define one multivariate Gaussian. The
long patches, say there are two or three per chain, represent expectation values
over many iterations. Small-scale features are averaged out, while the center of
gravity is preserved. Thus the initial output components from one group are very
similar, and the hierarchical clustering shifts and shrinks them to fit. It is
possible that some of the components are assigned zero weight, but due to the
initial similarity, very few, and usually zero, components ``die'' during the
hierarchical clustering.  Hence the chosen value of $K_g$, and thus $K$, is
preserved, which is the desired behavior.

In conclusion, let $n_g$ denote the number of chain groups, then the initial
mixture of output components for hierarchical clustering consists of $K = K_g
\cdot n_g$ components. Note that $K_g$ is required input from the user, but
$n_g$ is determined automatically as a function of the critical $R$ value
$R_{c}$, a parameter that requires only moderate tuning. The initialization is
described in pseudo code in \refalg{alg:hc-initialization}.

\begin{algorithm}[htb]
  \begin{algorithmic}
    \Require $k$ chains with $N_{\rm MCMC}$ samples
    \Require number of components per group $K_g $
    \State Start with empty initial mixture density $q^0_{\rm HC}(\cdot)$
    \State Discard the first $a \cdot N_{\rm MCMC}$ samples for burn-in
    \State Group the chains for given $R_c$
    \State $n_g \gets$ number of groups
    \ForAll{groups}
    \State $k_g \gets$ number of chains in group
    \If {$K_g < k_g$}
    \State Merge chains into one long chain
    \State $k_g \gets 1$
    \EndIf
    \State $\vecn \gets $ minimal lexicographic partition of $K_g$ into $k_g$ parts
    \ForAll {chains in group}
    \State Partition into $n_i$ patches
    \ForAll {patches in chain}
    \State Compute sample mean $\vecmu$ and covariance $\matsig$
    \State Add one component $\mathcal{N}(\cdot \| \vecmu, \matsig)$ to $q^0_{\rm HC}(\cdot)$
    \EndFor
    \EndFor
    \EndFor
    \State $q^0_{\rm HC}(\cdot) \gets$ Assign equal weight $\frac{1}{n_g K_g}$ to every component
  \end{algorithmic}
  \caption{\label{alg:hc-initialization}
    Initialization of the output components for hierarchical clustering.
    We use $a = 0.2$ and $R_c = 1.1 \dots 1.5$.
  }
\end{algorithm}

\subsubsection{PMC run}\label{sec:pmc-run}

The idea of adaptive importance sampling is to iteratively update the proposal
function $q$ to match the target density as closely as possible. In each step,
regular importance sampling is carried out, and adaptation is performed with an
expectation-maximization algorithm on mixture densities composed of Gaussian or
Student's t distributions \cite{Cappe:2004, cappe_adaptive_2008}. The proposal
in step $t$ is a mixture density
\begin{equation}
  \label{eq:mixture-density-step}
  q^t(\vecth) = \sum_{j=1}^K \alpha_j^t q^t_j(\vecth|\boldsymbol{\xi}^t_j),
  \quad q^t_j \in \{\mathcal{N}, \mathcal{T}_{\nu}\} .
\end{equation}
$q^t_j$ is a single multivariate component whose parameters are collectively
denoted by $\boldsymbol{\xi}^t_j = \left(\vecmu_j^t, \matsig_j^t\right)$. For
the Student's t case, $q^t_j = \mathcal{T}_{\nu}$, $\nu$ is the degree of
freedom. The set of (normalized) component weights is denoted by $\{ \alpha_j^t:
j=1,\dots,K\}$. Note that the component type, $\mathcal{N}$ or
$\mathcal{T}_{\nu}$ (including $\nu$), is fixed throughout a PMC run. The
$\mathcal{T}_{\nu}$ components may be preferable if the target has degeneracies
or fat tails.

The goal in each update step is to reduce the Kullback-Leibler divergence $\kl(P
\doubbar q)$ towards the minimum value of 0 at $q=P$. The general problem of
optimizing the $\kl$ functional is intractable. It is therefore necessary to
reduce the complexity to an ordinary parameter optimization problem by fixing
$q$ to the form \eqref{eq:mixture-density-step} and optimizing over $\left\{
  \left(\alpha_j, \boldsymbol{\xi}_j\right): j=1,\dots, K \right \} $. We want
to remark that in the basic formulation of \cite{cappe_adaptive_2008}, the
parameter $\nu$ is held fixed, but it could be updated along with the other
parameters through 1D numerical optimization~\cite{Hoogerheide:2011}.

For the Gaussian and Student's t case, the updated values $\alpha^{t+1}_j$ and
$\boldsymbol{\xi}^{t+1}_j$ are known, relatively simple-to-evaluate expressions
\cite{cappe_adaptive_2008} of $q^t$ and the importance samples
$\left\lbrace(\vecth_i^t, w_i^t): i = 1 \dots N\right\rbrace$ with importance
weights $w_i^t=P(\vecth_i^t) / q^t(\vecth_i^t)$.
It is important to stress again that PMC depends crucially on the initial proposal
$q^0$, because the updates tend toward only a \emph{local} minimum of $\kl$.

Two useful quantities to determine when PMC updates become unnecessary because
$q$ is ``close enough'' to $P$ are the \emph{perplexity}
\perpl{}~\cite{Wraith:2009if} and the \emph{effective sample size}
\ess~\cite{Liu:1995}. We normalize such that $\perpl, \ess \in [0,1]$, where the
optimal value of 1 is obtained for $P \propto q$, and thus $w_i = w_j \ \forall
i,j$. While $\mathcal{P}$ is sensitive rather to the mean of the distribution of
the importance weights, \ess{} is a function of their variance. In case of
successful PMC updates, $\perpl$ rises monotonically until reaching a plateau
(see for example \cite[Fig. 6]{Wraith:2009if}). As discussed in detail
in~\cite[Chapter 4.3]{Beaujean:2012}, \ess{} is sensitive to outliers; i.e.,
samples with a weight much larger than the average. Outliers cause \ess{} to
bounce up and down from one update to another
(cf. \cite[Fig. 4.7]{Beaujean:2012}), and render the \ess{} less robust as a
convergence criterion. As these outliers seem to be inevitable in higher
dimensions ($d \gtrsim 25$), we focus on \perpl{} as the sole convergence
criterion.  We stop PMC updates when the relative increase of \perpl{} is less
than \SI{5}{\%} in consecutive steps. A value of $\perpl > 0.9$ is attained only
in simple low-dimensional problems. The PMC algorithm in abstract form,
including our stoppage rule, is summarized in \refalg{alg:pmc-stoppage}.

\begin{algorithm}[htb]
  \begin{algorithmic}
    \Require number of samples $N$, initial proposal $q^0$
    \State converged $\gets$ false, $t \gets 0$ \Comment{Initialization}
    \While {$(t < t_{max}) \wedge (\neg \mbox{converged})$} \Comment{Update loop}
    \State draw $N$ samples $\vecth_i^t$ from $q^t$ and compute importance weights $w_i^t$ 
    \If {$t \geq t_{min} \wedge \left|\frac{\mathcal{P}^t -\mathcal{P}^{t-1}}{\mathcal{P}^t}\right| < \varepsilon $ }
       \State converged $\gets$ true
    \EndIf
    \State $q^{t+1} \gets $ update proposal based on  $q^t$ and $\left\lbrace(\vecth_i^t, w_i^t): i = 1 \dots N\right\rbrace$
    \State $t \gets t+1$
    \EndWhile
    \If {converged} \Comment{Final step}
    \State draw $N_{\rm final}$ samples from $q_{\rm final}$ and compute their importance weights
    \EndIf
  \end{algorithmic}
  \caption{\label{alg:pmc-stoppage} The generic PMC algorithm. We use
    $t_{min}=1$, $t_{max}=20$, $\varepsilon = \num{0.05}$.}
\end{algorithm}

The result of the first two stages of the algorithm, the MCMC prerun and the
hierarchical clustering, is a Gaussian mixture density $q_{\rm HC}$. Na\"ively,
we would set the initial proposal $q^{0}=q_{\rm HC}$, and start mapping the
target density with PMC. However, a number of considerations have to be taken
into account. We use Gaussians in the first two stages because the hierarchical
clustering is then particularly fast and simple to implement. But we do not
expect the chain patches turned into Gaussians to approximate the target density
with the highest precision. In particular, many realistic problems have thicker
tails, and are more accurately described by a Student's t mixture. In fact, a
much more involved hierarchical clustering for Student's t exists
\cite{Attar:2009}, but we don't expect it to reduce the number of PMC
updates. The sole purpose of $q_{\rm HC}$ is to cover the support of the target
with some accuracy, and the actual adaptation is left to the PMC update
algorithm. In the end, we only use the samples drawn from the adapted PMC
proposal for inference.  We therefore consider it appropriate to perform two
modifications to go from $q_{\rm HC}$ to $q^0$.

First, all component weights are set equal to balance the effect of an unequal
number of chains in each group. The weights are adjusted properly in the first
PMC update, so components are discarded if their target probability mass is
low, and not because few chains visited them.

Second, if a Student's t mixture is believed to yield a better representation of
the target, we create a ``clone'' of $q_{\rm HC}$ where each Gaussian component is
replaced by a Student's t component with identical location and scale
parameter. The degree of freedom, $\nu$, is the same for all components, and
currently has to be chosen a-priori by the user in the PMC approach.  Its
optimal value in the update is not known in closed form
\cite{cappe_adaptive_2008}. However, as noted in \cite{Hoogerheide:2011}, $\nu$
can be obtained from one-dimensional root finding. This is one source of future
improvement, as guessing the proper value of $\nu$ is not easy. In low
dimensions, the difference is usually small, but for large $d$, the impact of
outliers due to underestimating the tails may be significant, especially in
plots of 2D marginal distributions (for an example plot, see
\cite[Fig. 4.10]{Beaujean:2012}).

Assuming that $q^0$, the initial proposal, is fixed, there is still an open
question before we can start PMC: how many samples $N$ to draw from the
proposal? In the derivation of the PMC update step, $N \to \infty$ is assumed,
and this guarantees a reduced Kullback-Leibler divergence
\cite{cappe_adaptive_2008}. Large $N$ ensures many samples from each component,
but increases the computational burden. If $N$ is too small, the updates may
render $q^{t+1}$ worse than $q^{t}$, and the PMC algorithm fails. A proper
choice of $N$ depends mostly on the dimensionality of the target density $d$;
for guidance, cf. the discussion in \refsec{sec:short-guide-param}. After all,
$N$ is a required input to PMC, and is not deduced from the target. A reliable,
quantitative rule to determine $N$ would be very desirable, but is not available
to us. We then attempt to ensure that every component is explored initially, so
the quantity of interest is $N_c$, the \emph{number of samples per component} in
the first step, whence $N = K \cdot N_c$. Once the component weights are
adjusted in the first update step, components that receive a very low relative
weight are discarded, or ``die''; i.e., the number of samples drawn from them is
so small that there is not enough information gained to perform another
update. In the reference implementation of PMC \cite{pmclib:2011} that we use
for the updating, the minimum number of samples per component is set at 20. We
stop the update process when the convergence criteria of
\refalg{alg:pmc-stoppage} are met, and collect the samples used for inference in
the final step. Note that we do not have to keep $N$ constant in every step; in
fact, we have experimented with reducing $N$ as $N = K_{live} \cdot N_c$, where
$K_{live}$ is the number of live components. But we often saw PMC fail in those
cases, as after a short number of steps, $K_{live} \to 1$ resulting in
$\mathcal{P} \to 0$. Therefore, we recommend using identical values $N$ in every
PMC update step for improved stability. For the accuracy of inference, a larger
number of samples, $N_{\rm final}$, is advisable in the final step. At any step,
$Z$ and its uncertainty are estimated from importance weights $\left\lbrace w_i:
  i = 1 \dots N\right\rbrace$ through the sample mean and variance as
\begin{equation}
  \label{eq:Z-estimate}
    \widehat{Z} = \frac{1}{N} \sum_{i=1}^{N} w_i \, , \quad
    \widehat{V[Z]} = \frac{1}{N(N-1)} \sum_{i=1}^{N} \left(w_i - \widehat{Z}\right)^2    .
  \end{equation}

\section{Short guide to parameter settings}\label{sec:short-guide-param}

At this point, we summarize the previous sections and provide guidance on
setting the various tunable parameters.  Crucial settings of the particular runs
are listed in \reftab{tab:pmc-test-runs}.

For the MCMC step, we use $k=10 \dots 50$ chains depending on the expected
number of target modes. For a simple unimodal distribution, a handful of chains
should suffice. The chains are run for $N_{\rm MCMC}= \num{10000}$ ($d \gtrsim
2$) -- \num{100000} ($d=42$) iterations with a Gaussian proposal, though
Student's t could be used as well. For simple problems with $d$ small and a
decent initial chain proposal, the minimum value of $N_{\rm MCMC}$ is on the
order of \num{1000}.  Discarding the initial \SI{20}{\%} for burn-in, we split
up the chains into patches of length $L=$ 50 -- 300, the exact value of $L$ is
not critical.

With regard to hierarchical clustering, we group chains according to the $R$
values, using a threshold value around $R_c=1.2$. For larger dimensions or
respectively smaller $N_{\rm MCMC}$, larger values up to 1.5 or even 2 can be
used. The number of components per group, $K_g$, ought to be $\gtrsim d$; the
bigger $K_g$, the more accuracy is obtained at the expense of more evaluations
of the target. The initial components arise from long patches of chains within a
group. Hierarchical clustering is stopped if the distance measure in two
consecutive steps is reduced by less than $\varepsilon_{min}=\num{e-4}$.

In the PMC step, we initially set all component weights equal. In most
applications, a Gaussian mixture has tails that are thinner than the target's
tails, so one can decide for a Student's t mixture with degree of freedom $\nu=2
\dots 15$. Good results were obtained with $N_c$ ranging from 200 ($d=2$) over
600 ($d=20$) to 2500 ($d=42$). Convergence is declared when the normalized
perplexity $\perpl$ is stable to within \SI{5}{\%} between two consecutive
steps. Jumps in the ESS hint at outliers caused by too few mixture components or
by a proposal whose tails are too thin. If the PMC updates ``kill'' more and
more components and reduce the perplexity, more initial components and a larger
sample size may help. Another improvement may be to slightly increase $N_{\rm
  MCMC}$ or $k$. If outliers have a dominant effect on the resulting marginal
distributions, the combined effect of smoothing with kernel density estimation
and outlier removal provides a partial remedy. After convergence, a final sample
size $N_{\rm final}$ of as small as \num{5000} is sufficient for an integral
estimate at the percent level when $\perpl \lesssim 1$ and $d$ small. For the
toughest problems where $\perpl$ remains low, a size ranging in the millions is
necessary for targets in $d \gtrsim 30$.

\section{Examples}\label{sec:examples}
As described in the introduction, few publicly available codes exist for the
solution of difficult analysis problems. One is Multinest~\cite{Feroz:2008xx},
and we us this to benchmark our new approach in the following examples.

\subsection{Gaussian shells}\label{sec:gaussian-shells}
For easy comparison with the Multinest package, we use the same Gaussian-shell
example discussed in \cite[Sec. 6.2]{Feroz:2008xx} of two well separated
hyperspheres with a Gaussian density profile. We define the likelihood as
\begin{align}
  \label{eq:shell}
  & L(\vecth) = \frac{1}{2} \shell(\vecth | \vecc_1, r, w) + \frac{1}{2} \shell(\vecth | \vecc_2, r, w),
  & \shell(\vecth | \vecc, r, w) = \frac{1}{\sqrt{2 \pi w^2}} \exp \left[ -\frac{\left(\left| \vecth - \vecc \right| - r \right)^2}{2 w^2} \right] .
\end{align}
The width $w=0.1$ is chosen small compared to the radius $r = 2$, emulating a
problem with a continuous spherical symmetry (degeneracy). The two shells are
well separated by a distance of 7 units and do not overlap. The shell centers
are placed at
\begin{equation}
  \label{eq:shell-center}
  \vecc_{1,2} = ( \pm 3.5, 0, \dots , 0) ,
\end{equation}
and uniform priors over a hypercube $\vecth \in [-6, 6]^d$ are assumed. An
accurate analytical approximation of the evidence is
\begin{equation}
  \label{eq:shell-evidence}
  Z = \frac{\sqrt{2} \pi^{(d-1)/2}}{\Gamma(d/2) (2 \rho_{max})^d w}
  \int_0^{\rho_{max}} \rmdx{\rho} \rho^{d-1} \exp \left( {-\frac{(\rho -r)^2}{2 w^2}} \right) ,
\end{equation}
where the integral is performed over a hypersphere of radius $\rho_{max}=6$
covering a single shell. For the case at hand, the contribution from the
likelihood in the region contained in the hypercube but not in the hypersphere
is negligible. Note that there is an extra factor of $1/2$ in our
definition \eqref{eq:shell} compared to \cite{Feroz:2008xx}.

To assess the algorithm's performance, we repeat the run 100 times with
different pseudo-random numbers in $d=2,10,20$ dimensions. The parameter
settings are listed in \reftab{tab:pmc-test-runs}. For comparison, we also run
Multinest 100 times with parameter settings as advocated in \cite{Feroz:2008xx}
with 1000 live points and desired sampling acceptance rate of 0.3. To allow an
easier comparison, we fix the number of samples in the final PMC step, $N_{\rm
  final}$, at the average number of samples that Multinest yields. Note that
this does not equal the total number of target evaluations, $N_{\rm total}$,
with either algorithm. The Multinest algorithm accepts samples only with a
certain rate $\varepsilon$ such that $N_{\rm total}= N_{\rm final} /
\varepsilon$.  For our algorithm, the samples at the MCMC stage and during the
$t_{\rm final}$ PMC updates have to be added such that
\begin{equation}
  \label{eq:pmc-n-total}
  N_{\rm total} = k N_{\rm MCMC} + t_{\rm final} K N_c + N_{\rm final} .
\end{equation}

\subsubsection*{Discussion}

\begin{figure}[t]
  \centering
  \ifpgf
  \input{evolution-2D.pgf}
  \else
  \includegraphics{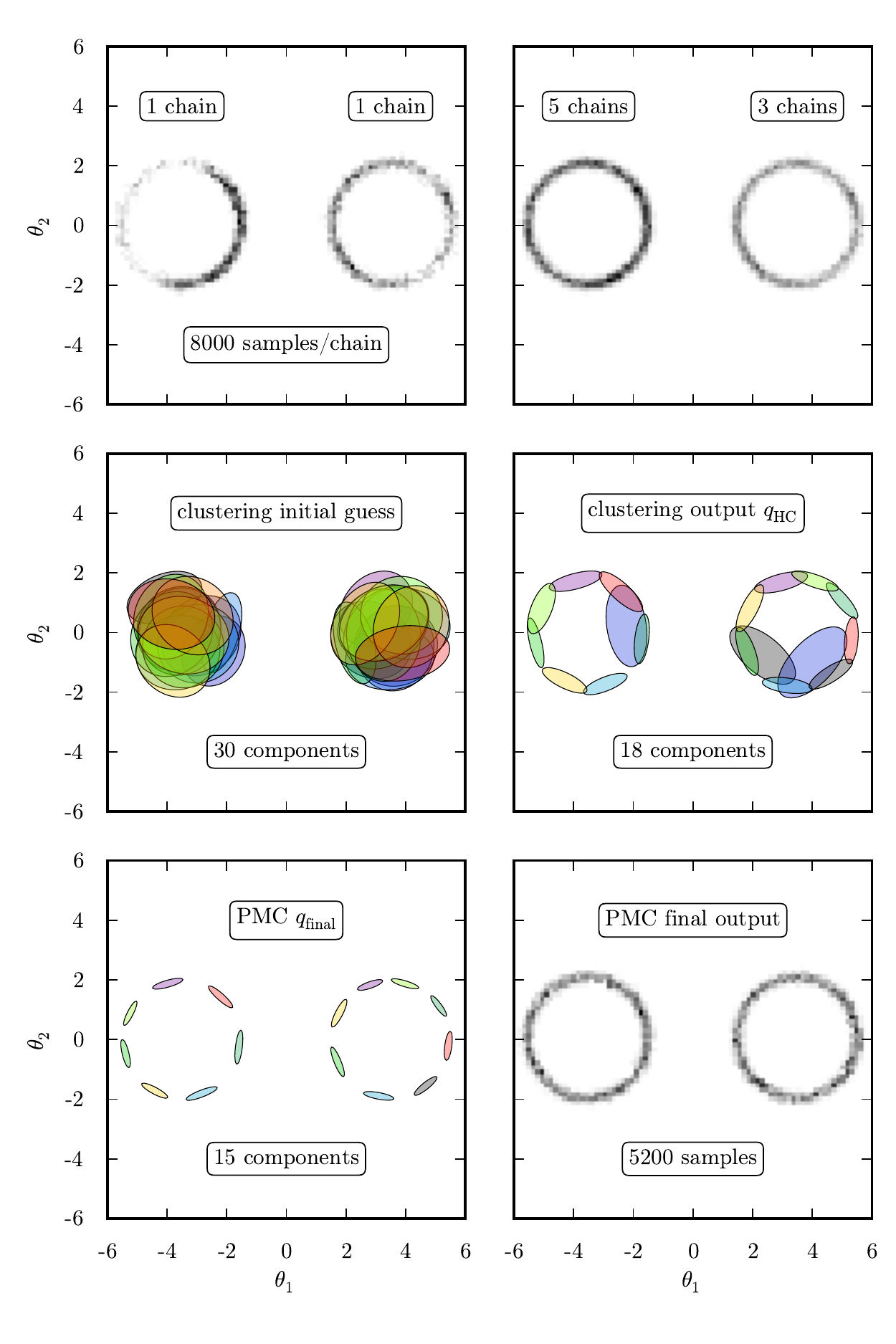}
  \fi

  \caption{\label{fig:shell-2D} Gaussian shell example in $d=2$. Histogram
    estimate of $P(\theta_1, \theta_2)$ from single chains (top left) and from
    combining all $k=8$ chains (top right). 1-$\sigma$ contours of the
    individual components of the Gaussian mixture density in hierarchical
    clustering for the initial guess (center left) and output (center right) and
    in the final PMC step (bottom left). The same color code is used in the
    latter two figures to identify components. Histogram estimate of
    $P(\theta_1, \theta_2)$ with samples of the final PMC step (bottom right).}
\end{figure}

We now comment on the performance of the algorithm in $d=2$, with relevant
settings and results summarized in \reftab{tab:pmc-test-runs} and
\reftab{tab:ex-perform}. In the MCMC step, individual chains may or may not
explore the entire region of one shell. Examples are shown in the top left panel
of \reffig{fig:shell-2D}. It is important that both shells are discovered, and
that the combination of chains covers both regions (\reffig{fig:shell-2D}, top
right), although the relative masses of the shells are incorrect because more
chains visit the left shell (5 versus 3). The large number of 640 chain patches
is used in hierarchical clustering to convert the initial guess with 30
components (\reffig{fig:shell-2D}, center left) into $q_{\rm HC}$
(\reffig{fig:shell-2D}, center right). The fact that approximately one third of
the components present in the initial guess are discarded during the clustering
demonstrates that the clustering may detect an unnecessary surplus of
components, a welcome feature. Nonetheless, there is a small number of
``outlier'' components in $q_{\rm HC}$, distinguished by the larger size and
location in the interior of a shell. These components are assigned a vanishing
weight during the 5 PMC updates needed to obtain the final proposal $q_{\rm
  final}$ (\reffig{fig:shell-2D}, bottom left). The final result
(\reffig{fig:shell-2D}, bottom right) accurately captures the two shells and
assigns equal probability mass to either shell, in contrast to the results
obtained from the combination of chains

In all of the 100 repetitions, the initialization is successful. PMC converges
quickly and determines the evidence to an accuracy of roughly \SI{1}{\%}
from only 5200 samples in the final step. Perplexity and \ess{} take on large
values, implying a good approximation of the target by the mixture density. Note
that the number of ``active'' components in the final step at about $K_{\rm
  final} = 17$ is significantly lower than the $K_g N_g = 30$ components
available at the beginning of hierarchical clustering. During the clustering, on
average 10 components are discarded, and only about 3 become inactive during the
PMC updates.  Extending to $d=10$ and $20$, the
evidence is again determined accurately at the percent level. We notice that the
fraction of runs, $f$, in which the correct evidence is contained in $[\estZ -
\estDelZ, \estZ + \estDelZ]$ diminishes slightly as $d$ increases, but remains
at a very reasonable value of $f=\SI{61}{\%}$ in $d=20$.

For $d>2$, somewhat surprisingly fewer PMC updates ($t_{\rm final} = 2-3$) are
needed than in two dimensions ($t_{\rm final} \approx 5$). This is likely due to
having relatively more components, and thus flexibility, available in the
proposal in $d=2$. For $d>2$, $\perpl$ and $\ess$ settle to lower values around
30--40~\%. This reduction in the maximum attainable $\perpl$ is common in higher
dimensions---a mild form of the ``curse of dimensionality''.  We note that both
shells are discovered, and all proposal components remain active throughout the
clustering and PMC updates. This demonstrates a successful adaptation due to the
good initialization from chains and hierarchical clustering.  By graphical
inspection, we verified that marginal distributions agree well with
expectations.

\begin{figure}[t]
  \centering
  \setlength{\relwidth}{\textwidth}
  \ifpgf
  \resizebox{\relwidth}{!}{\input{analyze-all.pgf}}
  \else
  \includegraphics[width=\columnwidth]{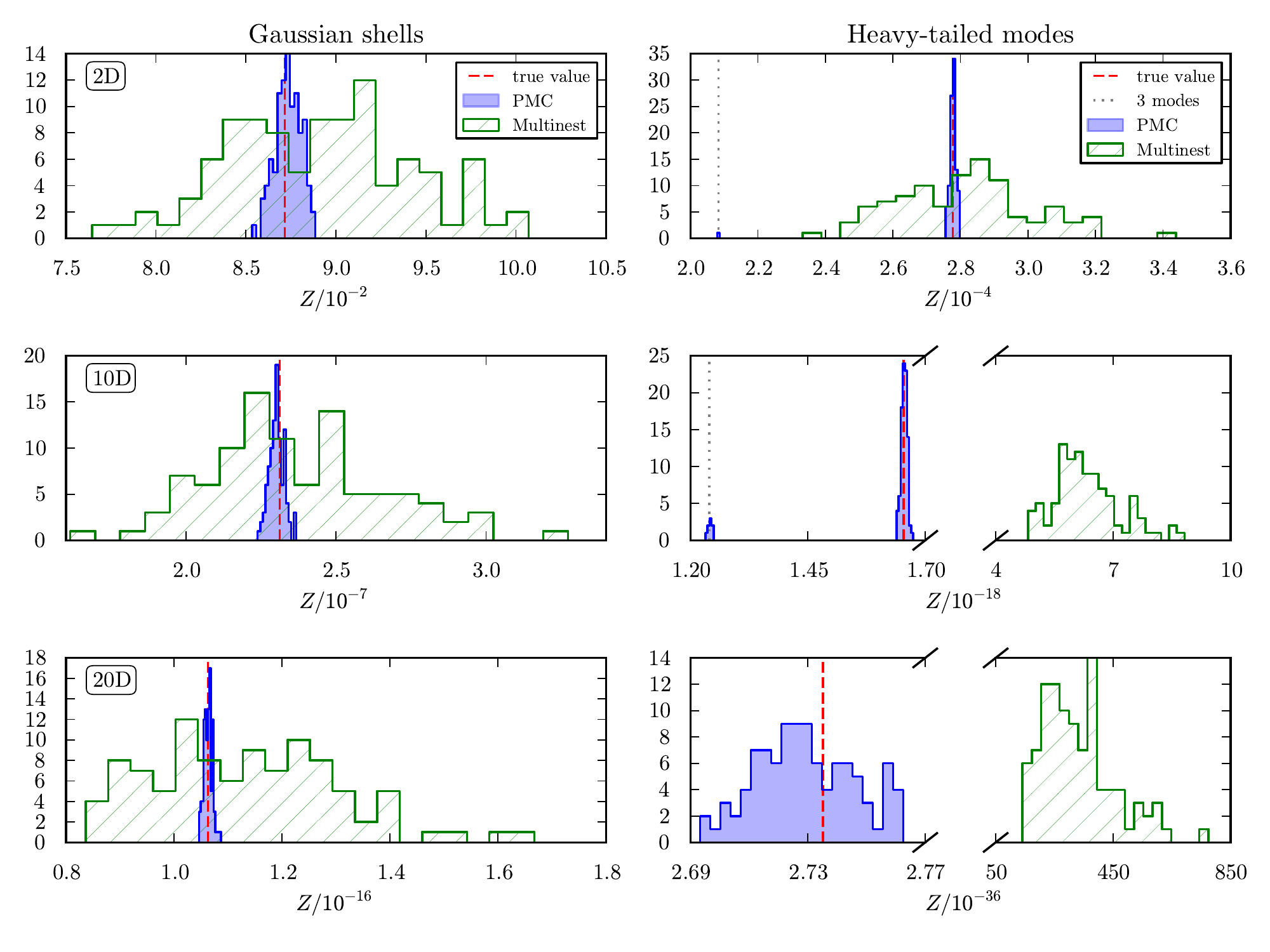}
  \fi
  \caption{\label{fig:Z-distributions} Evidence estimates of 100 runs of PMC
    with MCMC initialization and Multinest for the Gaussian shell (left column)
    and heavy tail (right column) examples in $d=2$ (top row), $d=10$ (center
    row), and $d=20$ (bottom row) dimensions. Note the very different scales
    used in each of the two lower plots on the right-hand side.  The dashed
    vertical line indicates the true evidence, whereas the dotted line gives the
    evidence when one of the modes is missed.}
\end{figure}

In the following, we want to compare directly how PMC and Multinest perform. In
general, both algorithms perform well; the two shells are properly explored, and
the correct evidence is found in the estimated interval in roughly 2/3 of the
runs (see \reffig{fig:Z-distributions} and \reftab{tab:ex-perform}). The
uncertainty estimate $\estDelZ$ is 5--20 times smaller with PMC for the same
number of samples considered, but at the expense of a larger $N_{\rm total}$. We
list the total running time and the average number of calls in
\reftab{tab:run-time}. To ensure a fair comparison, we did not use any parallel
evaluations, even though this is one of PMC's main strengths. Furthermore, we
replaced Multinest's slow default output to text files by output to the binary
HDF5 format~\cite{hdf5:2013}---the same format we use to store chains and PMC
output. In this example, the target density is quick to evaluate, hence the
majority of time is spent on updating the proposal in the PMC case, or on
clustering and finding a new point in Multinest.

As shown in \reftab{tab:run-time}, Multinest requires less times to run in all
considered cases and exhibits lower $N_{\rm total}$. While the exact numbers
depend on the algorithms' parameter settings, we observe that Multinest is ahead
on simpler problems, but our algorithm becomes competitive in higher dimensions
as Multinest's acceptance rate drops and $N_{\rm total}$ increases only moderately
with $d$ for our algorithm. We repeat that the timings were done in serial
execution, but PMC can easily use a large number of cores simultaneously and
hence is preferable for very costly target functions.

When it comes to integration precision, PMC is to be preferred, as it is able to
determine the evidence at an accuracy of better than one percent even in
$d=20$. In case the precision is considered too low \emph{after} the final step,
it is straightforward to load the stored final proposal from disk to sample more
points until reaching the desired precision with the usual $1/\sqrt{N_{\rm
    final}}$ scaling. In contrast, Multinest's precision is significantly lower,
even by a factor of roughly 20 in $d=20$, and cannot simply be improved by
continuing the run because of its serial nature of samples ordered by likelihood
value. For both algorithms, the relative estimated uncertainty on Z averaged
over the 100 repetitions, $E\left[\frac{\estDelZ}{\estZ}\right]$, agrees well
with the relative spread of the distribution of evidence estimates,
$\frac{\sigma[\estZ]}{E[\estZ]}$. The agreement is good to the last significant
digit given in \reftab{tab:ex-perform} for PMC, confirming the usefulness of the
uncertainty estimate and matching the near-Gaussian shape of the distributions
of $\estZ$ shown in \reffig{fig:Z-distributions}.

\begin{table}[htb]
  \centering
    \begin{tabular}{cccccccccccccc}
      \toprule
      &$d$&$k$&$N_{\rm MCMC}$ & $L$ &$\nu$&$K_g$&$N_c$&$N_{\rm final}$&$K_{\rm final}$&$t_{\rm final}$&$\perpl / \%$&$\ess / \%$ \\
      \multirow{3}{1em}{\begin{sideways} Shells\end{sideways}}
      &2  & 8 &  \num{10000} & 100 & -   & 15 & 200 & \num{5200}   & 17.3 & 5.02 & 75 & 51\\
      &10 & 8 &  \num{20000} & 100 & -   & 15 & 400 & \num{18000}  & 30 & 2 & 35 & 30\\
      &20 & 8 &  \num{20000} & 200 & -   & 25 & 600 & \num{40000}  & 50 & 2.9 & 43 & 36\\
      \midrule
      \multirow{3}{1em}{\begin{sideways} Tails\end{sideways}}
      &2  & 20 & \num{10000} & 100 & 12  & 5  & 200 & \num{6700}  &  19.95 & 1.4 & 95 & 93\\
      &10 & 20 & \num{20000} & 100 & 12  & 15 & 400 & \num{30000} &  58.8 & 2.2 & 81 & 71\\
      &20 & 20 & \num{20000} & 200 & 12  & 25 & 600 & \num{54000} &  99.91 & 3.5 & 56 & 34\\
      \bottomrule
    \end{tabular}
    \caption{\label{tab:pmc-test-runs}
      Settings and results of MCMC + PMC runs for the examples in $d$  dimensions.
      Each of the $k$ chains is run for $N_{\rm MCMC}$ iterations split into patches of length $L$. $\nu$ is the degree of freedom of each component
      in each $\mathcal{T}$ mixture proposal density, a missing value represents a $\mathcal{N}$ mixture.
      $K_g$ is the number of mixture components per group of chains, and $N_c$ is
      the number of samples per component drawn during the first PMC update step, while $N_{\rm final}$ is the fixed number of samples from all components in the final step.
      $K_{\rm final}$ is the number of active components after $t_{\rm final}$ PMC updates in the final step,
      in which $\perpl$ and $\ess$ characterize the quality of the adaptation of the proposal to the target. $K_{\rm final}$, $t_{\rm final}$, $\perpl$, and $\ess$ are averaged over 100 runs with the following common settings. During  MCMC, the Gaussian local random walk proposal function
      is updated after $N_{\rm update} = \num{200}$ iterations
      in $d=2$ and $N_{\rm update} = \num{500}$ iterations in $d> \num{2}$.
      Chains are grouped according to a critical $R$ value of $R_c=\num{1.2}$, and the first \SI{20}{\%} iterations are discarded for burn-in.
    }
\end{table}%

\begin{table}[htb]
  \centering
    \begin{tabular}{ccccccccccc}
      \toprule
      \multicolumn{3}{c}{}&\multicolumn{4}{l}{PMC}&\multicolumn{4}{l}{Multinest}\\
      &$d$&$Z$&$E[\estZ]$&$\frac{\sigma[\estZ]}{E[\estZ]}$ &$E\left[\frac{\estDelZ}{\estZ}\right]$&$f/\%$&$E[\estZ]$&$\frac{\sigma[\estZ]}{E[\estZ]}$&$E\left[\frac{\estDelZ}{\estZ}\right]$&$f/\%$\\
      \tabvsptop
      \multirow{3}{*}{\begin{sideways} Shells \end{sideways}}
      &2 &\num{8.726e-2}&\num{8.73e-2}&\num{0.008}& \num{0.009} & 69 & \num{8.9e-2} & \num{0.06} & $^{+0.052}_{-0.050}$ & 57\\
      \tabvsptop
      &10 &\num{2.304e-7} &\num{2.30e-7}&\num{0.011}& \num{0.012} & 71 & \num{2.4e-7} & \num{0.13} & $^{+0.13}_{-0.12}$ & 68\\
      \tabvsptop
      &20 &\num{1.064e-16}&\num{1.06e-16}&\num{0.007}& \num{0.007} & 61 & \num{1.1e-16} & \num{0.15} & $^{+0.21}_{-0.17}$ & 80\\

      \midrule

      \multirow{3}{*}{\begin{sideways} Tails\end{sideways}}
      &2 & \num{2.778e-4} & \num{2.78e-4}&\num{0.003}& \num{0.003} & 72 & \num{2.8e-4} & \num{0.07} & $^{+0.064}_{-0.060}$ & 63\\
      \tabvsptop
      &10 & \num{1.654e-18}&\num{1.66e-18}&\num{0.004}& \num{0.004} & 63 & \num{6.3e-18} & \num{0.13} & $^{+0.17}_{-0.14}$ & 0\\
      \tabvsptop
      &20 &\num{2.735e-36}&\num{2.73e-36}&\num{0.006}& \num{0.006} & 61 & \num{3.3e-34} & \num{0.37} & $^{+0.24}_{-0.19}$ & 0\\
      \bottomrule
    \end{tabular}
    \caption{
      Performance metrics of the 100 runs  for the examples in $d$  dimensions.
      $Z$ is the true evidence value.
      $E[\cdot]$ and $\sigma[\cdot]$ denote the sample mean and respectively the square root
      of the sample variance across all runs,
      whereas $\estZ$ and  $\estDelZ$ denote the evidence and respective uncertainty estimate
      from a single run according to \eqref{eq:Z-estimate}.
      $f$ is the fraction of runs in which $Z$ is contained in $[\estZ - \estDelZ, \estZ + \estDelZ]$.
      For the tail data in $d=2,10$, we use only the runs covering all four modes.
      Multinest's uncertainty estimate is transformed from the log scale to the linear scale
      and thus becomes asymmetric.
    }
  \label{tab:ex-perform}
\end{table}

\begin{table}[htb]
  \centering
  \begin{tabular}{ccccccc}
    \toprule
    &&\multicolumn{2}{l}{MCMC+PMC}&\multicolumn{3}{l}{Multinest}\\
    &$d$ & $t / s$ & $N_{\rm total}$ & $t/s$ & $N_{\rm total}$ & $\varepsilon / \%$\\

    \multirow{3}{1em}{\begin{sideways} Shells\end{sideways}}
    &2 & 1.6 & \num{105000} & 0.95 & \num{8300} & 63\\
    &10 & 5.4 & \num{202000} & 2.9 & \num{50000} & 36\\
    &20 & 45 & \num{274000} & 12.3 & \num{208000} & 19\\
    \midrule
    \multirow{3}{1em}{\begin{sideways} Tails\end{sideways}}
    &2 & 2.9 & \num{212300} & 0.4 & \num{18000} & 38\\
    &10 & 23 & \num{482800} & 13.7 & \num{269000} & 10\\
    &20 & 166 & \num{628000} & 108 & \num{5091000} & 1\\
    \bottomrule
  \end{tabular}
  \caption{Typical values of run time $t$, total number of target density evaluations $N_{\rm total}$,
    and acceptance rate $\varepsilon$ (Multinest only)
    solving the Gaussian shell (upper half) and heavy-tailed mode (lower half) examples
    on a single core of an Intel i7 920 clocked at \SI{2.67}{GHz}.}
  \label{tab:run-time}
\end{table}

\subsection{Heavy tails}\label{sec:heavy-tails}

The second example has four well separated maxima, each with the same shape and
probability mass. Such a distribution arises naturally in many high-energy
physics analyses in which the underlying physics model has discrete symmetries
in the parameters of interest. If the symmetry is not exact, it is of great
interest to accurately determine the relative masses of the maxima. We define
individual maxima as products of Gaussian and $\LogGamma$ 1D distributions. The
$\LogGamma$ distribution~\cite{Crooks:2010} is an asymmetric heavy-tailed
distribution with a location, scale, and shape parameter. In this form, the
example is a simplified yet hard-to-solve version of the posterior appearing in
our motivating analysis~\cite{Beaujean:2012uj}.

\begin{figure}[t]
  \centering
  \ifpgf
  \input{evolution-tail-20D.pgf}
  \else
  \includegraphics{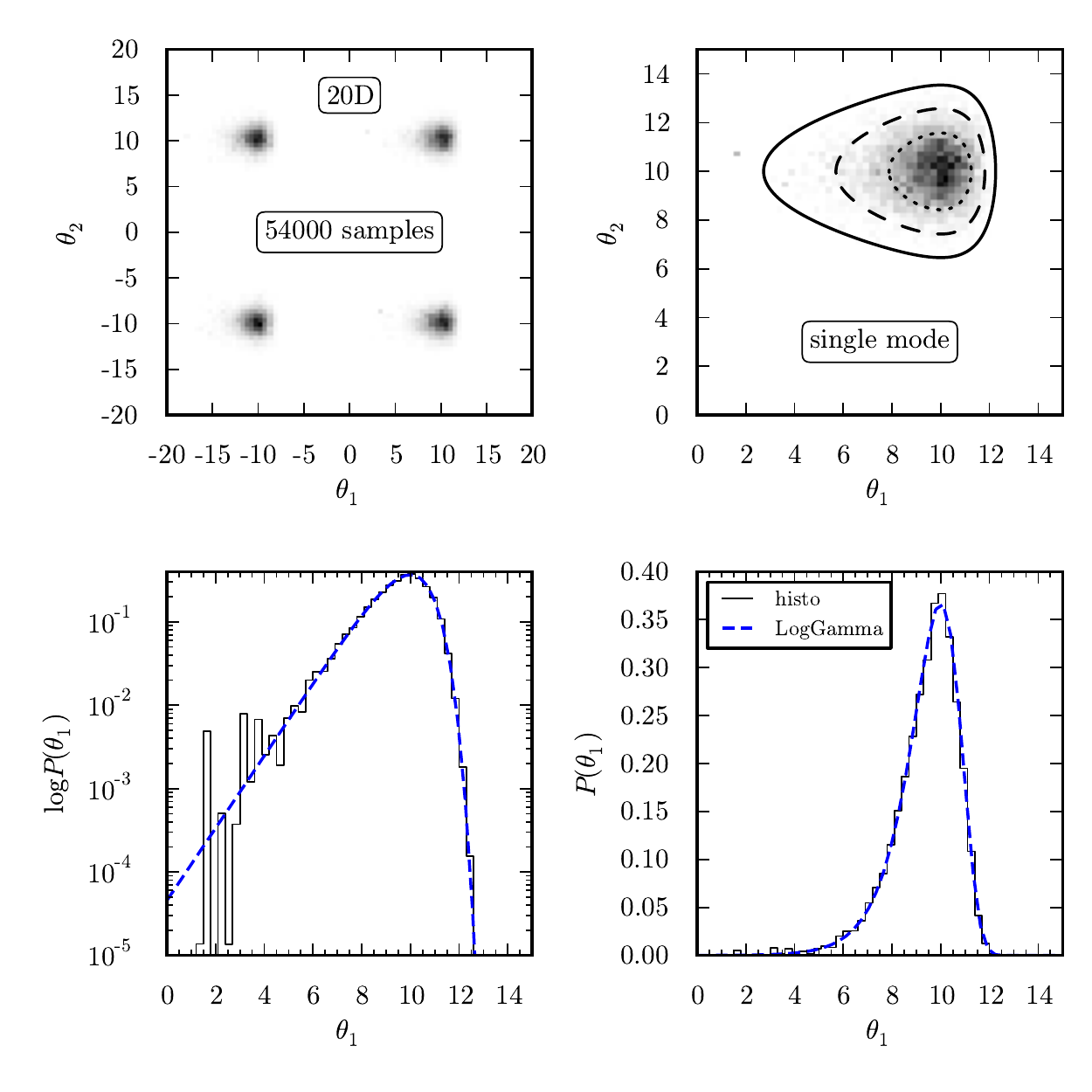}
  \fi
  \caption{\label{fig:tail-marginals} Heavy-tailed mode example in
    $d=20$. Histogram estimate of $P(\theta_1, \theta_2)$ for one PMC run with
    all four modes (top left). Zoom-in on a single mode around $(\theta_1,
    \theta_2)= (10, 10)$ with overlaid 1-, 2-, and 3-$\sigma$ contours obtained
    from integration on a fine grid (top right). The corresponding 1D marginal
    $P(\theta_1)$ on the log (bottom left) and linear (bottom right) scale is
    shown together with the true target $P(\theta_1) = \LogGamma(\theta_1 \| 10,
    1,1)$ overlaid.}
\end{figure}

For simplicity, we begin in $d=2$ dimensions where $L(\vecth) =
L(\theta_1,\theta_2) = L(\theta_1) \cdot L(\theta_2)$ (see top left panel of
\reffig{fig:tail-marginals}).  $L(\theta_1)$ is a mixture of two
$\LogGamma$ components with maxima at $\theta_1 = \pm 10$ with unit scale and
unit shape parameter. Similarly, $L(\theta_2)$ is a mixture of standard normal
distributions centered around $\theta_2 = \pm 10$:
\begin{align}
  \label{eq:ex-density}
    L(\theta_1)  &= 0.5 \LogGamma(\theta_1|10,1, 1)  + 0.5  \LogGamma(\theta_1|-10, 1, 1) \nonumber\\
    L(\theta_2)  &= 0.5 \mathcal{N}(\theta_2|10, 1 ) + 0.5 \mathcal{N}(\theta_2|-10,1 ) .
\end{align}

\noindent
To explore the effect of higher dimensions, we augment the 2D target density
with an equal number of $\LogGamma$ and Gaussian distributions. Throughout, we
assume uniform priors on $\vecth \in [-30, 30]^d$. Specifically, the target likelihood is
\begin{equation}
  \label{eq:multi-dim-example}
  L(\vecth) = \prod_{i=1}^{d} L(\theta_i) ,
\end{equation}
where the distribution in the first two dimensions is given by
\eqref{eq:ex-density}, and
\begin{align}
  \label{eq:extra-dim}
  L(\theta_i)   =
  \begin{cases}
    \LogGamma(\theta_i|10, 1, 1), \;\; 3 \le i \le \frac{d+2}{2} \\
    \mathcal{N}(\theta_i \| 10,1 ), \;\;  \frac{d+2}{2} < i \le d \; .
  \end{cases}
\end{align}
Since $L(\vecth)$ is normalized to unity, the evidence in $d$ dimensions is
given by the prior normalization as $Z(d) = 60^{-d}$. There are four modes due
to the first two dimensions; the extra dimensions do not add any further
modes. Hence, we perform $R$-value grouping as well as clustering in Multinest
only in the first two dimensions. The settings for MCMC and the initialization
are summarized in the lower left half of \reftab{tab:pmc-test-runs}. In contrast
to the Gaussian-shell example, we now use more chains ($k=20$) to increase the
chances to discover all modes, vary $K_g$ more strongly with $d$, and choose a
$\mathcal{T}$ mixture with $\nu = 12$ degrees of freedom in response to the
heavy tails of $\LogGamma$.

\subsubsection*{Discussion}

As with the Gaussian shells, we repeat the analysis 100 times with both MCMC+PMC
and Multinest. In the lower half of \reftab{tab:pmc-test-runs}, we list the
convergence properties of our algorithm. In all probed dimensions, essentially
all components remain active, convergence is reached after less than five steps
even in $d=20$, and perplexity reaches high levels. In summary, the
initialization procedure works well.

The results of the evidence calculations are shown in \reftab{tab:ex-perform}
and \reffig{fig:Z-distributions}. For PMC, both the estimated and the actual
accuracy are around an excellent value of \SI{0.5}{\%}. Note a common pitfall of
the Markov chain approach apparent in the bimodal distribution for $d=2$ and
$d=10$ in \reffig{fig:Z-distributions}. In one ($d=2$), respectively eight
($d=10$), of the runs, only three of the four modes are discovered, hence the
evidence is off by \SI{25}{\%} and there is no way for the algorithm to know
something is missing.  In an actual data analysis, one could gain extra
knowledge about the target, say from repeated mode finding, to learn about the
existence and location of various modes. Seeding the chains in different modes
would make the MCMC step both more reliable and more efficient. However, we use
this example both to show how well the algorithm performs with no such extra
knowledge given---just a fairly large number of $k=20$ chains---and to alert
unsuspecting users. Future developments should remedy this issue in a more
robust fashion; some ideas are discussed in \refsec{sec:outlook}. Focusing on
the runs in which all four modes are found, the fraction $f$ of runs for which
the true value is in $[\estZ - \estDelZ, \estZ + \estDelZ]$ gradually decreases
to \SI{61}{\%} in $d=20$.

Marginal distributions estimated from the PMC accurately approximate the target
density as shown in \reffig{fig:tail-marginals}, where we show 1D and 2D
marginals in $d=20$ for one example run. For $d \gtrsim 20$, one starts to
observe outliers. A mild outlier is visible in \reffig{fig:tail-marginals}
near $(\theta_1, \theta_2) = (2,11)$. Outliers are less of a problem in 1D
marginals; an example is shown in logarithmic and linear scale in
\reffig{fig:tail-marginals}. However, we stress that it is a common problem that
a single outlier can dominate a 2D marginal in $d \gtrsim 30$ if the proposal
density does not perfectly match the target. This is another manifestation of
the curse of dimensionality that plagues importance sampling methods in
general. Some remedies include smoothing of the marginal and filtering
outliers~\cite{Beaujean:2012}.

For comparison, we also ran Multinest with \num{1000} live points and the
desired acceptance rate set to 0.3 in all dimensions; cf.~\cite{Feroz:2008xx}
for a description of Multinest's parameters. Regarding mode finding, Multinest
is very robust as it discovers all four modes in every run. In $d=2$, Multinest
performs reasonably well, with an acceptance rate of \SI{38}{\%} and estimated
accuracy of \SI{7}{\%} with about a factor of 10 fewer calls to the target.  For
$d=10$, Multinest's acceptance rate reduces to \SI{10}{\%}, and the evidence is
overestimated\footnote{During the final stages of preparing this article, we
  were made aware of an ongoing effort by the Multinest authors to significantly
  improve the integration accuracy in the upcoming version 3.0.} by at least a
factor of 2.5 in all runs, the coverage dropping to zero. Similarly in $d=20$,
the evidence estimate is too large by a factor of at least 40, with $f=0$
despite an estimated uncertainty of roughly \SI{20}{\%}. The probability mass of
an individual mode ranges from 10 -- 58 \% compared to the correct value
\SI{25}{\%}. However, the 1D marginals agree well with expectations, and
furthermore there is no problem with outliers in Multinest, presumably because
the weight associated with each sample is based on a stochastic estimate of the
prior mass and is thus intrinsically smoothed.

The CPU usage is listed in \reftab{tab:run-time}. While Multinest roughly needs
a factor of 13 \emph{fewer} target evaluations in $d=2$, it requires eight times
\emph{more} in $d=20$ due its low acceptance rate of \SI{1}{\%}. Nonetheless,
PMC takes about \SI{60}{\%} longer for this simple target; this extra times is
spent almost exclusively in the proposal updates due the large number of 100
components.  Ideas for faster updates are discussed below.

\section{Outlook}\label{sec:outlook}

The proposed initialization of PMC with Markov chains and hierarchical
clustering performs well in the above examples. However, there are still
numerous improvements to make en route to the ideal black-box sampler.

\subsubsection*{Initialization}

The two most important aspects of the algorithm design to us are
\emph{correctness} and \emph{speed}. Regarding the former, it is crucial to
ensure that the algorithm leads to samples from the target; i.e., we need to
improve the automatic detection of \emph{all} regions of parameter space that
contribute. Within our framework, this could be achieved by choosing the chains'
starting points more cleverly, perhaps based on a preliminary sampling
step. This should lead to a reduction of the necessary number of chains, $k$,
for the majority of problems that have at most a handful of separated maxima.

Concerning speed, we consider reducing the overall execution time, and also
reducing the number of parameters steering the algorithm.  The latter helps in
two ways.  First, it requires time to assign a good value to each parameter,
either at run-time or by the user performing repeated trials. Second, a user may
inadvertently make a poor choice with adverse effects on the performance. The
most important parameter in this regard is $K$, the number of mixture
components.  A promising alternative to hierarchical clustering is the
\emph{variational Bayes} approach described in \cite{Bruneau:2010}, in which the
``best'' number of components is computed along with the positions and
covariances of the reduced mixture's components. We could obviate the patch
length $L$ by giving up slicing the chains into patches if we instead did the
clustering at the level of individual chain samples.

More fundamentally, one could eliminate the MCMC prerun entirely in favor of a
large number of samples from the prior or a uniform distribution on the
parameter space. Two advantages are that the samples can be computed with
massive parallelization and that potentially fewer samples are required by
avoiding the redundancy of multiple chains in the same region. This path is
followed in~\cite{Cornuet:2012} and shown to work for unimodal problems up to
$d=20$.  But by giving up the MCMC{} prerun, we suspect there is a greater
chance that suppressed modes and degeneracies are missed or poorly captured in
high-dimensional problems. In addition, it proved useful for validation purposes
to compare the marginal distributions from MCMC{} and PMC{} for qualitative
agreement. If a region is visible in the MCMC{} but not in the PMC{} output,
either PMC{} failed, or that region contains negligible probability mass, which
can be verified with the samples' target values in that region.  We expect only
minor improvements when replacing the Gaussian clustering with the considerably
more involved Student's t clustering to obtain a Student's t mixture proposal
density from the chain patches~\cite{Attar:2009}.

\subsubsection*{PMC}

Apart from the initialization, there are more general directions to further
enhance PMC. Our examples suggest that, given a good initial proposal,
importance sampling works well up to $d \approx 30$, and in fact we successfully
sampled from a single mode of the heavy-tailed example in $d=42$, but problems
with outliers appear already for $d \gtrsim 20$. To a certain degree, outliers
can be reduced by more mixture components, an adjustment of the Student's t
degree of freedom $\nu$, and more samples per component. We presented guidance
how to manually adjust these parameters, but an automatic adjustment is highly
preferred. $\nu$ can be determined in each PMC update by a 1D numerical solution
of Eq. (16) in \cite{Hoogerheide:2011}, making an informed user guess for $\nu$
obsolete. Providing even larger flexibility, individual components may then even
have different values of $\nu$. The soft limit of $d \approx 40$
(\cite{Hoogerheide:2011} noted a maximum of $d=35$ in their applications) is due
not only to outliers but also to the excessive time of updating the proposal,
whereas the MCMC initialization still works well.

It is conceivable that marginal distributions are less affected by outliers if
the proposal function of the final PMC{} step, $q_{\rm final}$, is used as a
global proposal in MCMC{}. Using $q_{\rm final}$ as a global proposal alone
would not solve that issue because the Metropolis-Hastings acceptance
probability~\cite{Hastings:1970} to accept a new point $\vecth_2$ given the
current point $\vecth_1$ is just the ratio of importance weights $w_2 / w_1$. An
outlier would be accepted with probability close to one, and the chain would
then be stuck for many iterations. Hence one needs a mixture of global and local
jumps for efficient sampling. We envision that---after appropriate scaling of
the covariance matrices---individual mixture components guide local jumps, and
the full proposal is used for global jumps.  Using a fixed proposal and assuming
rapid mixing due to the global jumps, massive parallelization is straightforward
and individual chains need to be run for fairly few iterations. Similar efforts,
though still involving a fairly large number of ad-hoc choices, are reported
in~\cite{Giordani:2010}.

The curse of dimensionality surfaces because the PMC update scales as
$\mathcal{O}\left(K N_c d^4 \right)$, where $K$ and $N_c$ have to be chosen
larger for increasing $d$. Due to the
Rao-Blackwellization~\cite{cappe_adaptive_2008}, each component density is
evaluated for every one of the $K N_c$ samples, and each such evaluation
requires a vector $\times$ matrix $\times$ vector product. Currently the update
is executed in serial code, but could be massively parallelized, and is easily
computed together with the target density. Another way to speed up would be to
partition the components such that for samples from the \nth{$j$} component only
the subset of other components needs to be considered that has a substantial
contribution. Ideally, the partitioning---similar to that employed in an
implementation of the fast Gauss transform~\cite{Morariu:2008}---could be done
at the component level before the weights are computed so each parallel process
would do only the minimum required evaluations.

\section{Conclusion}\label{sec:conclusion}

A new method is introduced to solve initialization difficulties in adaptive
importance sampling. The method was initially developed in the
context of a global fit to large data
sets~\cite{Beaujean:2012uj,Beaujean:2012}. It uses a combination of Markov
chains, hierarchical clustering~\cite{Goldberger:2004}, and population Monte
Carlo~\cite{Cappe:2004,cappe_adaptive_2008}. Our method was compared to a
publicly available implementation of nested sampling~\cite{Feroz:2008xx} for
examples with multimodal posterior distributions in up to 20 dimensions, and found
to perform well. The evidence was more accurate than that from nested sampling
and the marginal distributions were found to reproduce the target
distribution. The algorithm is amenable to massive parallelization.

The main development of this work consists in providing a reliable
initialization of adaptive importance sampling, allowing it to converge in very
few steps. While some tuning of parameters is still necessary, we consider this
to be an important step toward a ``black-box''sampling algorithm.

\section*{Acknowledgments}

The authors thank Danny van Dyk and Christoph Bobeth for numerous inspiring
discussions during the early development of the algorithm. We are grateful to
Martin Kilbinger for hints on using his PMClib package, and to Farhan Feroz for
guidance on Multinest. F.~Beaujean wishes to thank the IMPRS for elementary
particle physics for the generous support of this work.





\bibliographystyle{elsarticle-num}
\bibliography{references}







\end{document}